\documentclass[aps,prb,twocolumn,groupedaddress,floatfix]{revtex4}
\usepackage{graphics,graphicx}
\usepackage{color}
\usepackage{ulem}
\usepackage{wrapfig}
\usepackage{amssymb,amsmath}
\usepackage{bm}

\newcommand{\vp}{{\bm{p}}}
\newcommand{\vQ}{{\bm{Q}}}
\newcommand{\be}{\begin{equation}}
\newcommand{\ee}{\end{equation}}
\newcommand{\beq}{\begin{equation}}
\newcommand{\eeq}{\end{equation}}
\newcommand{\bea}{\begin{eqnarray}}
\newcommand{\eea}{\end{eqnarray}}
\newcommand{\et}{{\it et al.}}
\begin{document}
\unitlength = 1mm
\title{
Reduced effect of impurities on the universal pairing scale in the cuprates}

\author{ 
A.B.~Vorontsov$^1$, 
Ar. Abanov$^2$, 
M.G.~Vavilov$^3$, and 
A.V.~Chubukov$^3$
}

\affiliation
{$^1$~Department of Physics, Montana State University, Bozeman, MT 59717, USA \\
$^2$ Department of Physics, Texas A\&M University, College Station, TX 77843, USA\\
$^3$~Department of Physics, University of Wisconsin, Madison, WI 53706, USA}
\date{\today}


\begin{abstract}
We consider the effect of non-magnetic impurities on the 
onset temperature $T^*$ for
the $d-$wave pairing in spin-fluctuation scenario for the cuprates.  We
analyze intermediate coupling regime when the magnetic correlation length
$\xi/a >1$ and the dimensionless coupling  $u$ is $O(1)$. In the clean limit,
$T^* \approx 0.02 v_f/a$ in this parameter range, and weakly depends on $\xi$ and $u$. 
We found numerically  that this universal pairing scale is also 
quite robust with respect to impurities: 
the scattering rate $\Gamma_{cr}$ needed to
bring $T^*$ down to zero is about 4 times larger than in weak coupling, in
good quantitative agreement with experiments. 
We provide analytical reasoning for this result.  
\end{abstract} 

\maketitle

{\it Introduction.}~~ 
The issue of the pairing glue in the cuprates is still one of the hottest
topics in the physics of strongly correlated electrons.  Many researchers
believe that the pairing glue is a spin fluctuation exchange, at least in the
overdoped and optimally doped cuprates. At such dopings correlations are
relatively strong, but not enough so to break a large normal state Fermi surface
(FS) into the hole and electron pockets.

Recently, Abanov \et\  analyzed the pairing problem in the cuprates
within spin-fluctuation scenario, under the  assumption that the FS 
in the normal state  is large.\cite{abanov08}  Within this scenario, the
onset of the pairing at $T= T^*$ marks the development of the pseudogap
phase with strong thermal fluctuations of the pairing gap, 
and the true superconductivity emerges at a smaller $T_c \leq T^*$, when
thermal fluctuations become weaker. These authors found a smooth crossover
between the limit when the interaction $U$ is smaller than $v_f/a$,  and the
pairing is confined to a vicinity of hot spots 
(points on the FS separated by $(\pi,\pi)$), 
and the limit of strong interaction, when the entire FS is
``hot'' ($v_f$ is the bare Fermi velocity as obtained in band theory, and $a$
is the interatomic spacing, $v_f/a \sim 1 eV$). The onset temperature for the
pairing, $T^*$, scales as $T^* \sim (v_f/a) u$ for $u = 3Ua/8\pi v_f \ll 1$, 
and as $T^* \sim (v_f/a) (1/u)$ 
at large $u$, and weakly depends on $\xi$ for $\xi >a$. 
For intermediate values of $u = O(1)$,
which are mostly relevant to the cuprates, $T^*$ 
goes through a shallow maximum and is approximately $0.02 v_f/a$. The same
pairing scale was obtained in FLEX calculations for the Hubbard
model~\cite{FLEX}, in two-particle self-consistent calculations~\cite{tp}, in
dynamical cluster approximation~\cite{dca}, and in cluster DMFT~\cite{DMFT}.
The good agreement between all these results is strong indication that $T^*
\sim 0.02 v_f/a \sim 200-250K$ is indeed the universal pairing scale in
optimally doped cuprates. At smaller dopings, this scenario breaks down
because of electron localization which gives rise to precursors to hole and
electron FS pockets already in the normal state. The pseudogap temperature
$T^*$ then becomes  a scale at which the system develops such precursors,
while the pairing emerges at a 
 smaller temperature due to interaction between electron pockets.    

The subject of this communication is the analysis of how the universal pairing
scale $T^*$ in optimally-doped cuprates is affected by non-magnetic impurities,
which are pair-breaking for unconventional superconductors.
In near-optimally doped cuprates, concentrations of dopants 
are quite substantial, and potential random scattering 
off dopants could significantly reduce $T^*$.
At weak-coupling, which in our case corresponds to small $u$
{\it and} $\xi \sim a$, non-magnetic impurities in a $d-$wave superconductor 
 suppress  $T^*$ in the same way as magnetic impurities in a BCS
superconductor, and $T^*$ is given by  Abrikosov-Gorkov
(AG) formula ~\cite{AG}
$\log{T^*_{0}/T^*} = \Psi (1/2 + \Gamma/2\pi T^*) - \Psi (1/2)$, 
(with $\Gamma/2$ instead of $\Gamma$ for an $s$-wave and magnetic impurities~\cite{balatsky06}), 
where $\Psi(x)$ is the di-Gamma function and $T^*_{0}$ is the
pairing temperature in the absence of impurities. 
The ratio of the critical value of the 
scattering, $\Gamma_{cr}: T^*(\Gamma_{cr})=0$, to $T^*_{0}$ is 
$\Gamma_{cr} / T^*_{0} = \pi/2e^{0.5772} \approx 0.88$.
 
The issue we address here both analytically and numerically is what is this
ratio when $u = O(1)$ and $\xi \gg a$, when the pairing problem involves
incoherent fermions and near-gapless dynamical bosons and is very different
from the $d-$wave version of the BCS theory. We find that in this situation the
ratio $\Gamma_{cr}/T^*_0$ is about 4 times larger than $0.88$, i.e., the
pairing is much less suppressed by  impurities than in the weak coupling.
This result is in agreement with the experiments which
observed~\cite{tolpygo96} a similar reduction of the slope of $T^*(\Gamma)$
compared with the AG formula. 

Another issue that we consider here is how impurity scattering affects the
angular dependence of the $d-$wave pairing gap.  For a clean system, Abanov \et\
have found~\cite{abanov08} that in the universal regime the form of the gap
$\Delta_p (\omega)$ is very close to $\cos p_x - \cos p_y$ 
for all frequencies. We show that the $\cos p_x - \cos p_y$  form
 holds in the presence of
impurities -- the angular dependence only slightly changes with $\Gamma$. The
implication is that both $T^*$ and the gap structure are 
robust towards impurities. 

The angular dependence of the gap, particularly in underdoped cuprates, has
been  the subject of intensive debates recently, and some ARPES data were
interpreted as evidence for strong deviations from the $\cos p_x - \cos p_y$
form.  We emphasize in this regard that the position of the maximum of the
spectral function $A_p (\omega)$ represents the pairing gap $\Delta_p (\omega)$
only deep in the superconducting state. At higher $T$, the position of the
maximum in $A_p (\omega)$  differs from $\Delta_p (\omega)$ because of damping
induced by scattering off thermal bosons.  In particular, even for a gap with a
perfect $\cos p_x - \cos p_y$ form a maximum in $A_p (\omega=0)$ is still
present in some neighborhood of a node (a Fermi arc).   In this regard, our
result that the gap keeps $\cos p_x - \cos p_y$ form even in the presence of
impurities agrees with ARPES data by Campuzano \et,\cite{shi09} who detected
this form at the lowest $T$.    

The effects of non-magnetic and magnetic impurities in superconductors with
unconventional order parameters
have been studied  for
high-$T_c$ cuprates,\cite{mp94,franz97,haran98,kulik99,graser07,hirschfeld09}
non-cuprate  superconductors,\cite{mh71,yoksan84,preosti96,golubov97} 
and most recently for the pnictides.\cite{vorontsov09,dolgov09} For the cuprates,
most of the  studies attributed a slow decrease of $T^*$ to the extended nature
of the impurity potential~\cite{haran98,kulik99,graser07}, 
but Monthoux and Pines~\cite{mp94} performed a numerical 
analysis of $T^*$ suppression in YBCO by 
non-magnetic Ni impurities 
and found that the initial slope of $T^*$ is
quite small even when impurities are point-like scatterers. Very recently,
Kemper \et~\cite{hirschfeld09} studied the effect of disorder using
dynamical cluster approximation and quantum Monte Carlo, and found that ordinary
pair-breaking by impurities is partly balanced by the impurity-induced
enhancement of spin correlations which increases the 
pairing interaction mediated by spin fluctuations  

Our result agree with Kemper \et~\cite{hirschfeld09} 
and also Graser \et~\cite{graser07} in that the origin of the flattening of $T^* (\Gamma)$ are
magnetic strong-correlation effects.  At the same time, we found that, in  the
universal regime,  $T^*$ very weakly depends on the spin correlation length
$\xi$. In our theory, softness of $T^*$ suppression compared to weak-coupling
AG theory is primarily associated with the strong frequency dependence of the
pairing interaction. 

{\it Theory.}~~
We follow earlier work~\cite{abanov08} and consider fermions
with a large FS and $d-$wave pairing mediated by overdamped spin fluctuations.
We add to earlier analysis an isotropic, elastic  scattering by  point-like
impurities. As customary for the pairing problem, we introduce normal and anomalous
Green's functions and self-energies and treat spin-fluctuation mediated pairing
within the Eliashberg theory, by keeping self-energies but neglecting vertex
corrections. For small and large $u$'s, this approximation can be rigorously
justified because vertex corrections are small in $u$ or $1/u$, respectively.
For $u = O(1)$, it can only be justified on the basis that vertex corrections
are small numerically. \cite{abanov03}

The set of equations includes 
fermionic and bosonic self-energies in the normal state, 
and the linearized equation for the $d-$wave
pairing vertex $\Phi^\chi_{\vp_f}(\omega_m)$ (Ref.\onlinecite{abanov08}) 
\bea
&&
\Sigma_{\vp_f}(\omega_m) =  \pi T^* \sum_{\omega_m'} \int d\vp_f'
\chi_{\vp_f-\vp_f'}^{\omega_m-\omega_m'} \; \mbox{sign}(\omega_m \omega_m') \,,
\qquad 
\label{eq:I1} 
\\ 
&&
\Phi^{\chi}_{\vp_f} (\omega_m) =
- \pi T^* \sum_{\omega_m'} \int d\vp_f' 
\frac{\chi_{\vp_f-\vp_f'}^{\omega_m-\omega_m'}\Phi_{\vp_f'}(\omega_m') }
{|\omega_m'| + \Gamma + \Sigma_{\vp_f'}(\omega_m') } 
 \,,
\label{eq:I2}
\\
&&
\Phi_{\vp_f} (\omega_m) =
\Phi^{\chi}_{\vp_f}(\omega_m) + \Gamma \int 
\frac{ d\vp_f' \; \Phi_{\vp^{'}_{f}} (\omega_m)}
{|\omega_m| + \Gamma + \Sigma_{\vp_f'}(\omega_m)} \,,  
\label{eq:I3}
\\ 
&&
\chi_{\vp_f-\vp_f'}^{\Delta \omega}  =
\frac{(ua/\pi)}{(a/\xi)^{2} + a^2 |\vp_f -\vp_f' - \vQ|^2 
+ |\Delta \omega|/\Omega} \,,
\label{chi}
\eea
where  $\Omega = 3v_f/(16 u a)$ and momenta $\vp_f$ in all formulas are
confined to the FS, because integration in the direction transverse to the FS
has been carried out. In distinction to Ref. \onlinecite{abanov08}  in
(\ref{eq:I2}), (\ref{eq:I3})  we also included impurity renormalization of
$\Phi^{\chi}_{\vp_f}$, and of 
Matsubara energies 
$\omega_m = \pi T^* (2m+1)$, 
 where $\Gamma = (n_i/\pi N_f) \sin^2\delta$ depends on
impurity concentration $n_i$, fermionic density of states $N_f$, and the
impurity potential $u_0$ via $\tan\delta = \pi u_0 N_f$. 

The set of equations for $\Phi$ and $\Sigma$ can be simplified in the usual way
by introducing mass renormalization factor $Z_{\vp_f}(\omega_m)$ and the pairing
gap $\Delta_{\vp_f} (\omega_m)$ via 
\be 
Z_{\vp_f}(\omega_m) = 
\frac{|\omega_m| + \Gamma + \Sigma_{\vp_f}(\omega_m)}{|\omega_m|} \,, 
\quad 
\Delta_{\vp_f}(\omega_m) = \frac{\Phi_{\vp_f}(\omega_m)}{Z_{\vp_f}(\omega_m)} \,. 
\label{s_1} 
\ee 
Due to $A_{1g}$ symmetry of $\Sigma_{\vp_f}$ and 
$B_{1g}$ symmetry of $\Phi^\chi_{\vp_f}$ the impurity renormalization 
of the pairing vertex vanishes, i.e., $\Phi_{\vp_f} = \Phi^\chi_{\vp_f}$. 
 Using (\ref{s_1}) we obtain from (\ref{eq:I2})  
\begin{widetext} 
\be
\sum_{\omega_m'} \int d\vp_f' \left[
\pi T \frac{\chi_{\vp_f-\vp_f'}^{\omega_m-\omega_m'}
}{|\omega_m'| |\omega_m|} 
+ \delta_{m m'} \delta_{\vp_f \vp_f'} \; \frac{Z(\omega_m,\vp_f)}{|\omega_m|} 
\right] \Delta(\omega_m',\vp_f')  = 0. 
\label{eq:ev}
\ee
\end{widetext} 
We wrote the gap equation as an eigenvalue problem by moving 
all terms to one side and symmetrizing the kernel 
with respect to $\omega_m,\omega_{m'}$.

\begin{figure}[t]
\centerline{\includegraphics[width=0.9\linewidth]{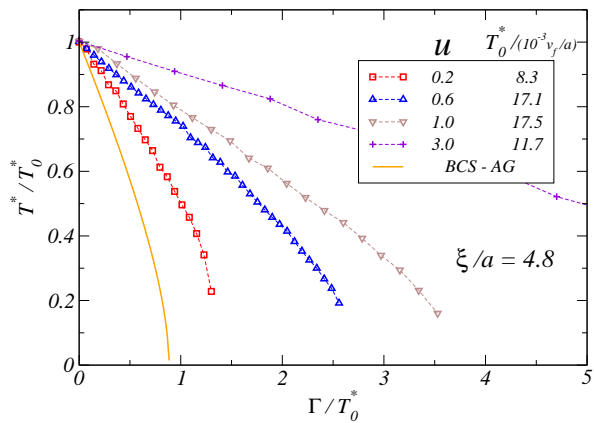}}
\caption{
    \label{fig1} (Color online) 
    The onset temperature for the pairing, $T^*$, vs. $\Gamma$ for different
    values of $u$. $T^*_0$ is the pairing temperature in 
    the clean limit. $v_f/a \sim 1 eV$ in the cuprates,
    hence $10^{-3} v_f/a \simeq 1 \, meV$.
    We set $\xi/a = 4.8$ for
    definiteness.  The solid line is AG-type result for a BCS $d-$wave
    superconductor, which in our case 
    corresponds to the limit $u \xi \ll 1$.
    The key result in this figure is a progressive increase with $u$ of the
    critical ratio $\Gamma_{cr}/T^*_0$, at which $T^*=0$. For $u = O(1)$, this
    ratio is about 4 times larger than in the BCS limit. The transition
    temperature is found with relative precision $10^{-2}$. The range of low
    $T$ requires special care because for any finite number of Matsubara
    points the curve $T^*(\Gamma)$ bends back towards the origin 
     producing a spurious second solution.} 
 
\end{figure}

{\it Numerical solution.}~~ 
This linearized gap equation is solved numerically by 
presenting the FS integral as a sum, 
varying $T$ and finding $T^*$ 
as the highest temperature where  Eq. (\ref{eq:ev}) is satisfied.
The result is presented in Fig.~\ref{fig1}.
We clearly see  a
strong increase of the  ratio $\Gamma_{cr}/T^*_0$ 
compared with a BCS $d-$wave superconductor. 
For $u \sim 1$, when $T^*_0$ as a function of $u$ has
a maximum  at about $0.02 v_f/a$, 
this ratio is nearly 4 times larger than
in the BCS limit. This result is in a good {\it quantitative} agreement with
the experiment in Ref.~\onlinecite{tolpygo96} 
and shows that the universal pairing scale
in the cuprates is  resistant 
to ordinary impurities.   

\begin{figure}[t]
\centerline{\includegraphics[width=0.9\linewidth]{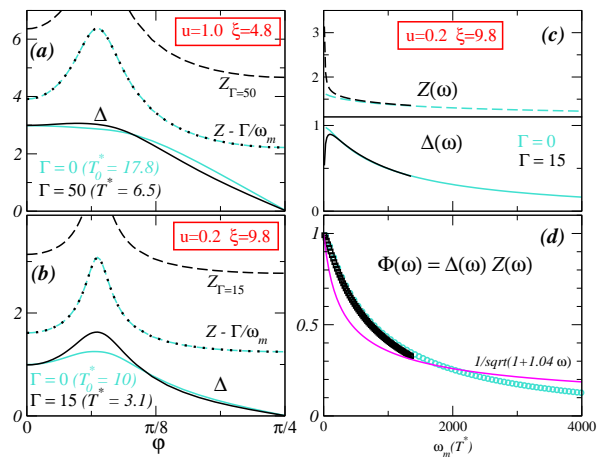}} \caption{
	\label{fig2} (Color online) Left: the angular dependence of the
	quasiparticle renormalization factor $Z_{\vp_f} (\omega_{m=0})$ and of
	the gap function $\Delta_{\vp_f} (\omega_0)$ for $u=1,\,\xi=4.8 a$ (a)
	and $u=0.2,\,\xi =9.8 a$ (b).  $T^*$, $\Gamma$ and $\omega_m$ are in
	units of $10^{-3} v_f/a$.  Observe that the angular dependences of the
	gap and Z (with subtracted constant shift due to $\Gamma$) changes very
	little between $\Gamma =0$(light lines) and $\Gamma \lesssim
	\Gamma_{cr}$(dark lines).  Panel (c): frequency dependences of
	$\varphi=0$ $\Delta_{\vp_f}(\omega_m)$ and $Z_{\vp_f} (\omega_m)$.  Panel
	(d): the comparison of the numerically obtained frequency dependence of
	$\Phi_{\varphi=0} (\omega_m)$ with the analytic solution of
	Eq.~(\ref{n_5}) in which momentum dependence of the gap near hot spot is
	neglected.  } \end{figure}

In Fig.~\ref{fig2} on the left we show
the angular dependences of the
quasiparticle renormalization factor 
 $Z_{\vp_f} (\omega_0) = 1 + [\Sigma_{\vp_f}(\omega_0)+\Gamma]/\omega_0$ 
and of the gap function $\Delta_{\vp_f} (\omega_0)$ 
for clean and dirty cases, for $\omega_0 = \pi T^*$ and different $u$ and $\xi$.  
In the clean case and $u=O(1)$ the angular dependence
is quite close to $\cos 2\phi$ 
(or $\cos p_x - \cos p_y$). 
We see that the effect of the impurities 
on the angular dependence of the gap is quite small, i.e., $\cos
2 \phi$ form is preserved in a dirty case. We verified that this holds for all
$\Gamma$ up to the critical value, and for all Matsubara frequencies. 
For completeness, in 
Fig.~\ref{fig2}(c) we show the frequency dependences of 
$Z_{\vp_f} (\omega_m)$ and $\Delta_{\vp_f} (\omega_m)$, 
and in panel (d) we present $\Phi_{\vp_f} (\omega_m)$ at $\varphi=0$ which we
will later compare with the analytical formula. 
  
{\it Analytical reasoning.}~~~ 
To understand the origin of the increase of 
$\Gamma_{cr}/T^*_0$  we analyze the  equation for the pairing vertex $\Phi_{\vp_f}
(\omega_m)$ analytically at $T=0$, i.e., we look for  a 
solution near $\Gamma_{cr}$. 
To do this, we make an approximation and neglect angular dependence of the gap
near a hot spot. The momentum integration along the FS then can be
carried out analytically, and the equation for the gap at a hot spot
becomes 1D integral equation in frequency only.  This equation is 
 more easy to analyze than the original $3D$ integral equation.  The
approximation of the gap function by a constant near hot spots can be
rigorously justified at small $u$ (corrections are higher powers of $u$),
but  remains qualitatively valid up to $u = O(1)$ (Ref. \onlinecite{abanov08}).
The expression for $T^*_0$ in this approximation has been obtained
earlier~\cite{abanov01} -- $T^*_0 \approx 0.13 u (v_f/a)$, with very
weak dependence on $\xi$ as long as $u \xi >1$. For small $u \xi$,
$T^*_0$ is described by a BCS formula.
   
Using $\Phi_{\vp_f + \vQ} (\omega_m) = - \Phi_{\vp_f} (\omega_m)$, dropping the
dependence of $\vp_f$ near a hot spot, integrating  over momentum in Eqs.
(\ref{eq:I1}) - (\ref{eq:I3}), and rescaling variables, we obtain after some
algebra the equation for $\Phi (\omega_m)$ 
at $T=0$ in the form 
\bea 
\Phi (x) &=&
\frac{\lambda}{2} \int_0^\infty dy \frac{\Phi (y)}{y + {\tilde \Gamma} +
\frac{2 \lambda y}{1 + \sqrt{4 \lambda^2 y +1}}} \times 
\nonumber 
\\  
&& \left(\frac{1}{\sqrt{4 \lambda^2 |y-x| + 1}} + 
\frac{1}{\sqrt{4 \lambda^2|y+x| + 1}}\right) \,,
\label{n_5} 
\eea 
where $x,y$ are frequencies in units of ${\bar \omega} = (3 u/4) v_f/a$, 
$\tilde{\Gamma} = \Gamma_{cr}/\bar\omega$, 
and $\lambda =(2 u \xi)$ 
(mass renormalization in the normal state is $1 + \lambda$). 

Weak coupling BCS limit corresponds to $\lambda \ll 1$. In this limit, 
$\Phi (x)$ becomes independent of $x$ (and $\Phi = \Delta$), the gap equation is
solved in the same way as in AG theory, and the  value of $\Gamma_{cr}/T^*_0 \approx 0.88$.  
We, however, are interested in the opposite limit, when $\lambda >1$.  
We analyzed  equation (\ref{n_5})  by normalizing $\Phi (x)$ to 
$\Phi (0) =1$, expanding at small $x$ and at large $x$ and extrapolating between
the two limits.  At small $x$, $\Phi = 1 - O(x)$, at large $x$, $\Phi (x)
\propto 1/\sqrt{x}$. 
We found that
$\Phi (x)$ is well approximated by $\Phi (x) = 1/\sqrt{1 + c x}$, 
where $ c$ is a constant which depends on $\lambda$ and $\tilde\Gamma$. 
The error 
is less than $2\%$ for all $x$ and for all $\lambda$ which we considered. We
 also checked that that this
solution is not an artefact: if we use this $\Phi (x)$ as an input and run
iterations, $\Phi (x)$ rapidly converges.  The critical value of ${\tilde
\Gamma}$ is then obtained by substituting this form back into (\ref{n_5}) and
solving  for $\Phi (0) =1$. 

Carrying out this procedure, we found that $c$ {\it increases} with increasing
$\lambda =u \xi$, i.e., when the correlation length increases and the pairing
problem becomes  more and more non-BCS, the gap function gets confined to
progressively smaller frequencies. The $c(\lambda)$ increases from $c(1)=0.66$, through 
$c(2) =1.04$ and $c(5) = 1.7$, to $c(\infty)=2.31$. 
In Fig.~\ref{fig2}(d) we compare our approximate analytic 
$\Phi (\omega_m)$ for $u=0.2,\,\xi = 9.8 a$
($\lambda \approx 2$) with the numerical $\Phi_{\vp_f} (\omega_m)$.
We see that the agreement is expectedly not prefect, but  generic
trends of the frequency dependence is captured by the approximate solution.   

Substituting $\Phi (x) = 1/\sqrt{1 + cx}$ back into (\ref{n_5}) and solving
 for $\Phi (0) =1$, we found that ${\tilde \Gamma}$
progressively increases as $\lambda$ gets larger, from $\lambda^{-2}
e^{-1/\lambda}$ at small $\lambda$ to $0.3$ for $\lambda =1$ and to $0.46$ for
$\lambda = \infty$. 

The  monotonic 
increase of the               
value of ${\tilde \Gamma}$ with increasing $\lambda$
 is a tricky effect. One could expect that the
confinement of $\Phi (x)$ to smaller $x$ as $\lambda$ increases 
 and the increase of the self-energy tend to reduce
${\tilde \Gamma}$ simply because typical frequencies get
smaller. However,
 as $\lambda$ increases, the interaction strength also increases, 
 and this tends to increase ${\tilde\Gamma}$ because ${\tilde \Gamma}$  
appears in the denominator  in the integral for
$\Phi (0)=1$, and larger  ${\tilde \Gamma}$ are required to balance the
increase of the interaction. We compared the two effects and found that
increase of the interaction overshadows other effects and is the origin
 of the growth of 
$\tilde\Gamma$  with increasing $\lambda$.

We next compared the growth   
of  
$\Gamma_{cr} = \bar\omega \tilde\Gamma(\lambda)$ 
and the grown of 
$T^*_0$. The latter also scales as $\bar\omega$ with 
 $\lambda$-dependent prefactor (Ref. \onlinecite{abanov01}). 
This prefactor  increases with increasing $\lambda$, 
but its $\lambda$-dependence is very weak: 
it changes by less than $5\%$ between $\lambda =1$ and $\lambda = \infty$. 
As a result, the $\lambda$-dependence of the ratio $\Gamma_{cr}/T^*_0$
predominantly comes from $\tilde\Gamma(\lambda)$, 
which, we remind, increases with $\lambda$.
 Inserting  the numbers, 
we find that the  ratio $\Gamma_{cr}/T^*_0$
becomes $2.0$ for $\lambda =1$; $2.37$ for $\lambda =2$; 
$2.47$ for $\lambda =5$, and $2.74$ for $\lambda = \infty$. 
The scale of the increase is quite
consistent with what we found numerically in Fig.~\ref{fig1} 
by solving the full 3D integral equation in momentum and frequency. 

{\it Summary.}~~ 
In this paper we considered the effect of
non-magnetic impurities on the onset temperature $T^*$ for the $d-$wave pairing 
in spin-fluctuation scenario for the cuprates.  Non-magnetic impurities are
pair-breaking for  $d-$wave superconductivity, and one should expect a
reduction of $T^*$ due to impurities. In weak-coupling, $T^*$ falls off 
rapidly, following the AG curve. 

We  analyzed the effect of impurities in the
intermediate coupling regime when the magnetic correlation length $\xi/a >1$, 
 the dimensionless coupling $u$ is $O(1)$, and 
the pairing problem 
is qualitatively different from BCS. 
In the clean limit, $T^*$ in this parameter range weakly depends 
on $\xi$ and $u$ and  is approximately $0.02 v_f/a$. 
We found that this universal pairing scale is quite robust with
respect to impurities: the critical value of the scattering 
rate $\Gamma_{cr}$ needed to bring $T^*$ down to zero is about 4 times larger 
than in the weak coupling.
This  implies that the slope of 
the initial reduction of $T^*$ is weaker by about the same factor 
than in the weak coupling. This reduction of the
slope agrees with the experiments~\cite{tolpygo96} and with earlier work by
Monthoux and Pines\cite{mp94} on  $T^*$ suppression in YBCO  due to
non-magnetic Ni impurities. We also analyzed the angular dependence of the
gap and found that it is little affected by impurities.

We thank I.Vekhter for useful discussions. Ar. A. is supported by 
NSF 0757992, and 
Welch Foundation (A-1678), A.V.Ch. is supported by
 NSF-DMR-0906953.  
Three of us (A.B.V., M.G.V. and A.V.Ch.) 
are thankful to the Aspen Center for Physics 
for hospitality.


\end{document}